\documentclass[12pt]{article}
\usepackage{latexsym}
\usepackage{amsmath}
\usepackage{amssymb}
\usepackage{verbatim}
\begin{document}
\begin{center}
\Large
\textbf{What Kind of a Friend Is Wigner's?}\\[0.5cm]
\large
\textbf{Louis Marchildon}\\[0.5cm]
\normalsize
D\'{e}partement de chimie, biochimie et physique,\\
Universit\'{e} du Qu\'{e}bec,
Trois-Rivi\`{e}res, Qc.\ Canada G9A 5H7\\[0.2cm]
(\verb+louis.marchildon@uqtr.ca+)\\
\end{center}
%
%
\begin{abstract}
Proietti \emph{et al.}\ (arXiv:1902.05080)
reported on an experiment designed to settle,
or at least to throw light upon, the paradox
of Wigner's friend. Without questioning the
rigor or ingenuity of the experimental
protocol, I argue that its relevance to the
paradox itself is rather limited.
\end{abstract}
%
\section{Wigner's Friend}
In a paper entitled ``Remarks on the
Mind-body Question''~\cite{wigner},
published in~1961, E. P. Wigner argued
that ``consciousness enters [quantum
mechanics] unavoidably and unalterably.''
In a process of observation or measurement,
the state vector (or wave function) changes,
according to Wigner, when an impression
enters into our consciousness.

This, Wigner saw at once, easily leads
to a contradiction. Let
\begin{equation}
\frac{1}{\sqrt{2}} (|h\rangle + |v\rangle)
\end{equation}
be the state vector of a
photon,\footnote{$|h\rangle$ and $|v\rangle$
represent horizontal and vertical polarization
states. The notation follows
Ref.~\cite{proietti}.} and let a `friend'~$F$
observe the photon polarization in the
$\{|h\rangle, |v\rangle\}$ basis. The friend's
consciousness then becomes~$|F_h\rangle$ if she
sees~$|h\rangle$ and $|F_v\rangle$ if she
sees~$|v\rangle$. But suppose the whole process
is initially hidden from Wigner. He should then
attribute to the photon-friend system the
state vector
\begin{equation}
\frac{1}{\sqrt{2}} (|h\rangle |F_h\rangle
+ |v\rangle |F_v\rangle) .
\label{corr}
\end{equation}
This state is, in principle, different
from $|h\rangle |F_h\rangle$, $|v\rangle
|F_v\rangle$ or any mixture thereof,
whence the contradiction.

Wigner realized that the contradiction could
be dissolved if consciousness were
attributed to him only. He pointed out,
however, that ``to deny the existence of the
consciousness of a friend to this extent
is surely an unnatural attitude, approaching
solipsism, and few people, in their hearts,
will go along with it.''
To him, the
contradiction is resolved by noting that
following his friend's observation, the
true state of the photon-friend system is
either $|h\rangle |F_h\rangle$ or $|v\rangle
|F_v\rangle$.

Wigner also pointed out that if $|F_h\rangle$
and $|F_v\rangle$, instead of representing
states of a conscious being, represent atomic
states that can be correlated with the photon's
polarization, the true photon-atom state is
indeed given by~(\ref{corr}).
%
\section{An Unconscious Friend\label{aif}}
Most current interpretations of quantum
mechanics make no appeal to consciousness
in the formalization of a measurement
process.\footnote{An exception is
QBism~\cite{fuchs}, where the state vector
represents an agent's information, or
betting preferences.} Let us look briefly
at three of them: the pilot wave, spontaneous
localization and many worlds. In each case,
we will let the conscious friend be replaced
by an inanimate instrument (also denoted
by~$F$) designed to measure photon
polarization.

In the pilot-wave approach~\cite{broglie,bohm},
each particle has a well-defined position at
all times and follows a deterministic
trajectory. Assuming that the photon-$F$
system is isolated, (\ref{corr}) represents
the true state vector, whether $F$ stands
for a macroscopic instrument or an atom.
In the first case, the configuration of the
particles making up the instrument will
quickly concentrate either in a region where
(in configuration space) $F_h$ is nonzero and
$F_v$ vanishes, or in a region where $F_v$ is
nonzero and $F_h$ vanishes. For all practical
purposes, subsequent evolution will proceed
as if (\ref{corr}) had only one
term.\footnote{If Wigner has technology
sufficiently advanced to monitor macroscopic
interference, both terms are still
relevant.} In the
case where $F$ stands for an atom, however,
$F_h$ and $F_v$ may not always have vanishing
intersection in configuration space, so that
both terms of~(\ref{corr}) must be kept.

With spontaneous localization~\cite{ghirardi},
what happens is different, but the end result
is essentially the same. If $F$ stands for
an atom, the number of elementary particles
involved is perhaps~100. The probability of
localization of one particle in the next
second is typically taken as $10^{-16}$.
Therefore, the probability of localization
of the atom during the course of measurement
(say 10$^3$\,s) is on the order of
\begin{equation}
100 \times 10^3\, \mbox{s} \times 10^{-16}\,
\mbox{s}^{-1} \approx 10^{-11} ,
\end{equation}
a very small number. Hence both terms
of~(\ref{corr}) must be kept. If $F$
stands for an instrument, however, perhaps
10$^{25}$ particles are involved. Localization
will then happen with probability close to~1
in less than a nanosecond. One of the terms
of~(\ref{corr}) essentially disappears.

In the many-worlds approach~\cite{everett},
both terms of~(\ref{corr}) must be kept at
all times. Let the splitting of `worlds'
occur with macroscopic systems
only.\footnote{For alternatives see
Ref.~\cite{marchildon}.} If $F$ stands for
an atom, there is no split, and both terms
of~(\ref{corr}) are relevant at all times.
But if $F$ stands for an instrument, the
result of the split is that each of the
two worlds is governed by only one term
of~(\ref{corr}) (but see footnote~3).

The upshot is that in all three interpretations
examined, Wigner will attribute state vector
(\ref{corr}) to the photon-$F$ system if $F$
stands for an atom, and only one term
of~(\ref{corr}) if $F$ stands for an instrument.
%
\section{Experimental Investigation}
Inspired by Brukner's no-go
theorem~\cite{brukner},
the experiment described by Proietti
\emph{et al.}~\cite{proietti} consists in
setting up a procedure which in the end
prepares four photons in the state
\begin{align}
|\Psi\rangle
&= \frac{1}{\sqrt{2}} \cos \frac{\pi}{8}
(|h\rangle_a |F_v\rangle_a |v\rangle_b |F_h\rangle_b
+ |v\rangle_a |F_h\rangle_a |h\rangle_b |F_v\rangle_b)
\notag\\
& \qquad + \frac{1}{\sqrt{2}} \sin \frac{\pi}{8}
(|h\rangle_a |F_v\rangle_a |h\rangle_b |F_v\rangle_b
- |v\rangle_a |F_h\rangle_a |v\rangle_b |F_h\rangle_b).
\label{state}
\end{align}
Indices $a$ and $b$ refer to Alice and Bob, each
of whom gets one photon of a pair originally
prepared in the entangled state
\begin{equation}
\frac{1}{\sqrt{2}} (|h\rangle_a |v\rangle_b
- |v\rangle_a |h\rangle_b ). 
\end{equation}
The experimental protocol correlates, on
Alice's side, the state $|h\rangle_a$ with
the state $|F_v\rangle_a$ and the state
$|v\rangle_a$ with the state $|F_h\rangle_a$
(and similarly on Bob's side). In this sense,
although $|F_v\rangle_a$, $|F_h\rangle_a$,
$|F_v\rangle_b$ and $|F_h\rangle_b$ are
one-photon states, they are formally
associated with `friends.'

Proietti \emph{et al.} also introduce
four observables $A_0$, $A_1$, $B_0$ and $B_1$,
the first two associated with Alice and the
last two with Bob. Alice's observables are
defined as
\begin{equation}
A_0 = \Bbb{I} \otimes
(|F_v\rangle_a \langle F_v|_a
- |F_h\rangle_a \langle F_h|_a)
\label{valfr}
\end{equation}
and
\begin{align}
A_1 &= \frac{1}{2} (|h\rangle_a |F_v\rangle_a
+ |v\rangle_a |F_h\rangle_a)
(\langle h|_a \langle F_v|_a
+ \langle v|_a \langle F_h|_a) \notag\\
& \qquad - \frac{1}{2} (|h\rangle_a |F_v\rangle_a
- |v\rangle_a |F_h\rangle_a)
(\langle h|_a \langle F_v|_a
- \langle v|_a \langle F_h|_a) , 
\label{valw}
\end{align}
with similar definitions on Bob's side. Operators
$A_0$ and $A_1$ act as the identity on Bob's
state space, and similarly with $B_0$ and $B_1$
on Alice's space. Thus $A_0$ and $A_1$ commute
with $B_0$ and $B_1$.

It is straightforward to show that
$A_0^2 = \Bbb{I} \otimes \Bbb{I}$ and that
\begin{equation}
A_1^2 = |h\rangle_a |F_v\rangle_a
\langle h|_a \langle F_v|_a + |v\rangle_a |F_h\rangle_a
\langle v|_a \langle F_h|_a ,
\end{equation}
with similar relations for $B_0^2$ and
$B_1^2$. From this we conclude that $A_0$ and
$B_0$ can only have eigenvalues $\pm 1$,
whereas $A_1$ and $B_1$ can only have
eigenvalues $0, \pm 1$. If we assume that
all these observables have simultaneous values
(or, we could say, are simultaneous elements
of reality), we easily check that
\begin{equation}
A_1 B_1 + A_1 B_0 + A_0 B_1 - A_0 B_0 \le 2 .
\end{equation}
This implies that if relevant quantities are
adequately sampled in experimental runs,
\begin{equation}
\langle A_1 B_1 \rangle + \langle A_1 B_0 \rangle
+ \langle A_0 B_1 \rangle
- \langle A_0 B_0 \rangle \le 2 .
\label{CHSH}
\end{equation}
Experimental results have revealed a
violation of~(\ref{CHSH}) by five standard
deviations~\cite{proietti}.
%
\section{Discussion}
The experiment shows that $A_0$, $A_1$,
$B_0$ and $B_1$ cannot have simultaneous
values. Since, however, the $A$'s commute
with the $B$'s, any~$A$ can have simultaneous
value with any~$B$. The upshot is that either
$A_0$ does not have simultaneous value with
$A_1$, or $B_0$ does not have simultaneous
value with $B_1$. From~(\ref{valfr}) we see
that the value of $A_0$ corresponds to
Alice's friend's information about the photon's
polarization, while from~(\ref{valw}) we see
that the value of $A_1$ corresponds to 
Alice's information about the superposition
of photon-friend states.

It is important to note that these results
are completely consistent with unitary quantum
mechanics. Since $A_0$ and $A_1$ do not
commute, they are not expected to have
simultaneous values.

Proietti \emph{et al.}~\cite{proietti} interpret
their result as an ``experimental rejection of
observer-independence.''
If one observer
registers a value, the other doesn't. To reach
this conclusion, they are careful to define an
observer as ``any physical system that can extract
information from another system by means of
some interaction, and store that information
in a physical memory.''
In this sense, even
a single photon can be an observer. Such a
definition is vastly different from what Wigner
had in mind when introducing his friend who,
as we saw, is a conscious being.\footnote{This
is also stressed in Ref.~\cite{bong},
in connection with another Wigner's friend
experiment.} Wigner
believed that replacing a conscious friend
by an atomic system would lead to behavior
consistent with unitary quantum mechanics.
Thus he would have expected the experimental
results just described. It would certainly be
interesting to investigate how these results
would evolve if the `observer' were made
more and more complex. As pointed out
in~\cite{proietti}, however, there are easier
ways to probe quantum mechanics at larger
scales.

Proietti \emph{et al.}\ also claim that
their result ``lends considerable strength to
interpretations of quantum theory already
set in an observer-dependent
framework\footnote{Refs.~\cite{fuchs}
and~\cite{rovelli}, for example.}
and demands for revision of those
which are not.''
But all interpretations
of quantum mechanics mentioned in
Sect.~\ref{aif} predict, for the system
they used, the experimental results
they obtained. From an empirical point
of view, therefore, these results provide
no support for an observer-dependent
framework over any commitment one already
has to such framework.

To conclude, the notion of observer
introduced in~\cite{proietti} is far removed
from the original notion of Wigner's friend.
And technically nice as they are, these
results provide no reason to prefer one
interpretation of quantum mechanics over
others.
%

%

\begin{thebibliography}{99}
%
\bibitem{wigner} E. P. Wigner,
 ``Remarks on the Mind-Body Question,''
in {\it The Scientist Speculates},
I. J. Good, ed.\ (William Heinemann, 1961),
pp.~284--302;
reprinted in E. P. Wigner,
{\it Symmetries and Reflexions}
(Ox Bow Press, 1979), pp.~171--184.
%
\bibitem{proietti} M. Proietti, A. Pickston,
F. Graffitti, P. Barrow, D. Kundys, C.~Branciard,
M. Ringbauer and A. Fedrizzi,
``Experimental Rejection of Observer-Independence
in the Quantum World,''
arXiv:1902.05080v1.
%
\bibitem{fuchs} C. A. Fuchs,
N. D. Mermin and R. Schack,
``An Introduction to QBism with an Application
to the Locality of Quantum Mechanics,''
{\it American Journal of Physics}
{\bf 82}, 749--754 (2014).
%
\bibitem{broglie} L. de Broglie,
``La m\'{e}canique ondulatoire et la structure
atomique de la mati\`{e}re et du rayonnement,''
{\it Le journal de physique et le radium}
{\bf 8}, 225--241 (1927).
%
\bibitem{bohm} D. Bohm,
``A Suggested Interpretation of the Quantum
Theory in Terms of `Hidden' Variables,
I and II,''
{\it Physical Review}
{\bf 85}, 166--193 (1952).
%
\bibitem{ghirardi} G. C. Ghirardi,
A. Rimini and T. Weber,
``Unified Dynamics for Microscopic and
Macroscopic Systems,''
{\it Physical Review D}
{\bf 34}, 470--491 (1986).
%
\bibitem{everett} H. Everett III,
``\,`Relative State' Formulation of
Quantum Mechanics,''
{\it Reviews of Modern Physics}
{\bf 29}, 454--462 (1957). 
%
\bibitem{marchildon} L. Marchildon,
``Multiplicity in Everett's Interpretation
of Quantum Mechanics,''
{\it Studies in History and Philosophy of
Modern Physics}
{\bf 52}, 274--284 (2015).
%
\bibitem{brukner} \v{C}. Brukner,
``A No-Go Theorem for Observer-Independent Facts,''
{\it Entropy} {\bf 20}, 350 (2018).
%
\bibitem{bong} K.-W. Bong, A. Utreras-Alarc\'{o}n,
F. Ghafari, Y.-C. Liang, N. Tischler,
E. G. Cavalcanti, G. J. Pryde and H. M. Wiseman,
``Testing the Reality of Wigner's Friend's
Experience,''
arXiv:1907.05607v2.
%
\bibitem{rovelli} C. Rovelli,
``Relational Quantum Mechanics,''
{\it International Journal of
Theoretical Physics}
{\bf 35}, 1637--1678 (1996).
%
\end{thebibliography}
\end{document}